\documentclass[conference]{IEEEtran}
\IEEEoverridecommandlockouts
\usepackage{cite}
\usepackage{amsmath,amssymb,amsfonts}
\usepackage{algorithmic}
\usepackage{graphicx}
\usepackage{textcomp}
\usepackage{xcolor}
\usepackage{xspace}
\usepackage{url} 
\usepackage{comment}

\usepackage[
  letterpaper,
  top=0.75in,
  bottom=1.04in,
  left=1.015in,
  right=1.015in
]{geometry}

\usepackage{tikz}

\usepackage{cite}
\usepackage{amsmath,amssymb,amsfonts}
\usepackage{algorithmic}
\usepackage{graphicx}
\usepackage{textcomp}
\usepackage{xcolor}
\usepackage{xcolor}
\usepackage{cleveref}
\usepackage{comment}
\usepackage{booktabs}
\usepackage{atbegshi}

\newcommand{\hide}[1]{} %

\def\BibTeX{{\rm B\kern-.05em{\sc i\kern-.025em b}\kern-.08em
    T\kern-.1667em\lower.7ex\hbox{E}\kern-.125emX}}
\begin{document}

\newcommand\submittedtext{%
\footnotesize \textbf{Preprint}. This work has been accepted to the IEEE International Workshop on Decentralized Physical Infrastructure Networks 2025, in conjunction with the IEEE International Conference on Blockchain and Cryptocurrency (ICBC 2025). © 2025 IEEE. Personal use of this material is permitted. Permission from IEEE must be obtained for all other uses, in any current or future media.}

\newcommand\submittednotice{%
\begin{tikzpicture}[remember picture,overlay]
\node[anchor=south,yshift=10pt] at (current page.south) {\fbox{\parbox{\dimexpr0.65\textwidth-\fboxsep-\fboxrule\relax}{\submittedtext}}};
\end{tikzpicture}%
}

\title{Trustworthy Decentralized Autonomous Machines: A New Paradigm in Automation Economy
}
\author{\IEEEauthorblockN{Fernando Castillo}
\IEEEauthorblockA{\textit{Technische Universität Berlin} \\
\textit{Berlin}, Germany \\
fc@ise.tu-berlin.de}
\and
\IEEEauthorblockN{Oscar Castillo}
\IEEEauthorblockA{\textit{Dobprotocol} \\
\textit{Delaware}, USA \\
oscar@dobprotocol.com}
\and
\IEEEauthorblockN{Eduardo Brito}
\IEEEauthorblockA{\textit{Cybernetica AS} \\
\textit{University of Tartu} \\
\textit{Tartu}, Estonia \\
eduardo.brito@cyber.ee}
\and
\IEEEauthorblockN{Simon Espinola}
\IEEEauthorblockA{\textit{Dobprotocol} \\
\textit{Delaware}, USA \\
simon@dobprotocol.com}

}

\maketitle
\submittednotice

\begin{abstract}
Decentralized Autonomous Machines (DAMs) represent a transformative paradigm in automation economy, integrating artificial intelligence (AI), blockchain technology, and Internet of Things (IoT) devices to create self-governing economic agents participating in Decentralized Physical Infrastructure Networks (DePIN). Capable of managing both digital and physical assets and unlike traditional Decentralized Autonomous Organizations (DAOs), DAMs extend autonomy into the physical world, enabling trustless systems for Real and Digital World Assets (RDWAs). In this paper, we explore the technological foundations, and challenges of DAMs and argue that DAMs are pivotal in transitioning from trust-based to trustless economic models, offering scalable, transparent, and equitable solutions for asset management. The integration of AI-driven decision-making, IoT-enabled operational autonomy, and blockchain-based governance allows DAMs to decentralize ownership, optimize resource allocation, and democratize access to economic opportunities. Therefore, in this research, we highlight the potential of DAMs to address inefficiencies in centralized systems, reduce wealth disparities, and foster a post-labor economy.%

\end{abstract}


\section{Introduction}
\label{sec:introduction}

Decentralized Physical Infrastructure Networks (DePIN) have emerged as a novel paradigm for building and managing real-world infrastructure using blockchain and cryptoeconomic incentives~\cite{lin2024decentralized,ballandies2023taxonomy}.
This model enables individuals and organizations to collectively operate physical systems – such as wireless networks, energy grids, and transportation services – while earning token rewards and ownership stakes for their contributions.
In recent years, the DePIN sector has grown rapidly, with approximately 300 projects with over 21M devices active already in 2025\footnote{https://depinscan.io/}, indicating broad interest across domains from IoT connectivity to energy and mobility.
Parallel to this, AI-driven automation is propelling a wave of autonomous systems in the physical world~\cite{xia2023towards}. Advances in artificial intelligence, sensors, and connectivity have enabled self-driving vehicles, smart factories, and intelligent devices that are transforming industries, ranging from transportation and manufacturing to agriculture and public services.

However, realizing the fusion of AI systems and DePIN is challenging, as decentralized networks and autonomous machines have largely evolved in isolation. Most autonomous systems today still rely on centralized control, which introduces single points of failure and limits transparency and adaptability in complex operations.

Building on this momentum, we introduce the concept of Decentralized Autonomous Machines (DAMs) as a unifying approach to fuse DePIN principles with AI capabilities. In our vision, a DAM leverages community-driven networks for distributed data, compute, and governance, while employing AI for real-time decision-making and adaptation, moving beyond traditional decentralized organizations~\cite{chiu2024depin,gong2024tokenised}. Such machines can operate as self-governing entities that overcome the limitations of centralized automation, offering greater transparency, resilience, and efficiency.

In this paper, we address the current gaps and explore the synergy between DePIN and AI, offering insights into developing a new generation of intelligent, decentralized machines. We investigate the theoretical framework, technical architecture, and socio-economic implications of DAMs. Additionally, we argue that DAMs are critical in transitioning from trust-based to trustless economic models, enabling scalable, transparent, and equitable solutions for managing Real and Digital World Assets (RDWAs). The integration of AI agents, IoT devices, and blockchain technologies turns DAMs into a novel approach to decentralizing ownership, optimizing resource allocation, and democratizing access to economic opportunities.

The remainder of this paper is structured as follows: the technological foundations of DAMs are introduced in \Cref{sec:relatedWork}, their role in transitioning to trustless economic models is discussed in \Cref{sec:model}, their technical realization in \Cref{sec:tech_realization}, their socioeconomic implications are introduced in \Cref{sec:Evaluation}, and the conclusion in \Cref{sec:Conclusion}.

\section{Driving Forces}
\label{sec:relatedWork}
The emergence of DAMs is rooted in the convergence of several technological paradigms, including DePIN, AI, blockchain, and IoT. This section examines developments that facilitate the advancement of DAMs.

\subsection{Decentralized Physical Infrastructure Networks }
DePIN represent a novel approach to managing real-world infrastructure through blockchain-based incentives and decentralized governance~\cite{pillai2024sok}. Projects such as Helium~\cite{dzhunev2022helium} and Filecoin~\cite{guidi2022evaluating} exemplify this model, where participants are incentivized to contribute to physical networks (e.g., wireless connectivity, storage) in exchange for token rewards. Decentralized incentive mechanisms, in the context of IoT, have been explored before~\cite{sizan2025evaluating}, which closely overlaps with recent DePIN models~\cite{lin2024decentralized, pillai2024sok, fan2023towards}.%

\subsection{Integration of AI, Blockchain, and IoT}
The integration of AI and blockchain within IoT systems is a pathway for enabling intelligent, secure, and autonomous operations. AI has already been employed to enhance the decision-making capabilities of IoT devices, allowing them to process data and respond to environmental changes in real-time~\cite{alahi2023integration}. Meanwhile, Blockchain has been used to ensure data integrity and secure transactions, providing a trustless framework for machine-to-machine (M2M) interactions~\cite{kumar2021tp2sf}. The intersection of these technologies has also been examined in the literature. For instance, Bothra et al.~\cite{bothra2023can} discuss how AI can optimize blockchain operations in IoT contexts through real-time data analytics and automated decision-making. %

\subsection{Automation Economy through Smart Contracts}
Smart contracts on blockchain platforms enable automated economic transactions without intermediaries. In the context of IoT, smart contracts facilitate direct economic interactions between devices, such as buying and selling services or data~\cite{suliman2019monetization}. DeFi platforms, as explored by Schär~\cite{schar2021decentralized}, exemplify this shift by using smart contracts to execute transparent and efficient financial transactions. %

\subsection{Case Studies and Prototypes}
There exist case studies and prototypes to illustrate the practical implementation of technologies involving DePIN and AI. For example, in~\cite{bendiab2023autonomous}, the use of blockchain in autonomous vehicle networks has been examined, addressing challenges in reliable and verifiable transactions. Another case is decentralized energy trading platforms  that demonstrate how peer-to-peer energy exchange can be managed in a decentralized manner~\cite{wongthongtham2021blockchain}.%

\subsection{Comparative Analysis and Future Directions}
A recent comparative analysis of 530 studies shows a progression from foundational blockchain-IoT integrations~\cite{saidu2025convergence} to more advanced applications involving AI and automation~\cite{ernst2019economics}. This evolution suggests future research directions, such as developing standardized governance protocols to explore M2M models across industries. 

\vspace{1em}
While these technological components form the foundation of DAMs, their transformative potential lies in how they may fundamentally alter economic relationships and value exchange mechanisms. This shift necessitates a deeper examination of the economic principles underlying decentralized systems.

\section{The Transition to Trustless Economic Models}
\label{sec:model}
The transition from trust-based to trustless frameworks marks a significant evolution in economic value creation and exchange. This section examines the limitations of centralized systems — with high costs, high opacity, and single points of failure — and explores how blockchain and AI enable trustless alternatives. %

\subsection{From Trust-Based to Trustless Systems}

Traditional economic systems rely on intermediaries (e.g., banks, brokerages) to facilitate trust, requiring participants to have faith in their honesty and efficiency. Trustless systems, powered by blockchain, a distributed ledger governed by immutable code, eliminate this need, as participants trust algorithms and consensus mechanisms rather than institutions~\cite{malik2019trustchain, zetzsche2020decentralized}. Decentralized finance (DeFi) exemplifies this shift, using smart contracts on public blockchains to execute transactions without intermediaries~\cite{zetzsche2020decentralized}. These transparent and automated protocols reduce costs, errors, and counter-party risk, enabling value exchange without a centralized authority.

\subsection{Inefficiencies of Centralized Finance and Asset Management}

Centralized systems suffer from several inefficiencies driving the move to trustless models:
\begin{itemize}
    \item \textbf{High Costs and Friction:} Intermediaries add fees and delays, whereas trustless DeFi platforms settle transactions instantly with lower costs~\cite{chen2020blockchain}.
    \item \textbf{Lack of Transparency:} Opaque processes limit verification, unlike decentralized systems offering full on-chain visibility~\cite{sedlmeir2022transparency}.
    \item \textbf{Single Points of Failure:} Centralized control risks catastrophic failures, while decentralized protocols remain resilient~\cite{zarrin2021blockchain}.
    \item \textbf{Limited Inclusivity:} Gatekeepers restrict access and innovation; decentralized networks lower barriers, fostering participation~\cite{murinde2022impact}.
    \item \textbf{Operational Rigidities:} Scaling centralized infrastructure is resource-intensive and slow, in contrast to blockchain-based systems, which can adapt through modifications to the consensus mechanism, data sharding, or Layer 2 solutions~\cite{zhou2020solutions}.
\end{itemize}

\subsection{AI-Powered Efficiency in Trustless Networks}

AI enhances trustless systems with automation and adaptability. Integrated with blockchain, AI-driven smart contracts can respond to real-time data (e.g. market prices), reducing reliance on intermediaries~\cite{dhar2024blockchain}. AI has also benefited from and contributed to more transparent operations by logging decisions on-chain, and has enhanced security by detecting fraud in DeFi or by optimizing DePIN resources~\cite{raval2021smart,luo2024ai}. Furthermore, blockchain provides AI agents with verifiable identities, enabling reliable M2M transactions in a trustless environment~\cite{gong2024tokenised}. %

\vspace{1em}
Nevertheless, the transformation of economic principles into functioning systems requires a new technical architecture that integrates physical operations, intelligent decision-making, and blockchain-based governance. We addressed this in the following section.

\section{Technical Realizations}
\label{sec:tech_realization}

This section presents the technical framework of DAMs, focusing on their layered architecture, key enabling technologies, and solutions for scalability and interoperability. This framework integrates IoT, AI, and blockchain to enable autonomous machine operations in decentralized, trustless systems.

\begin{figure}[h]
    \centering
    \includegraphics[width=0.8\columnwidth]{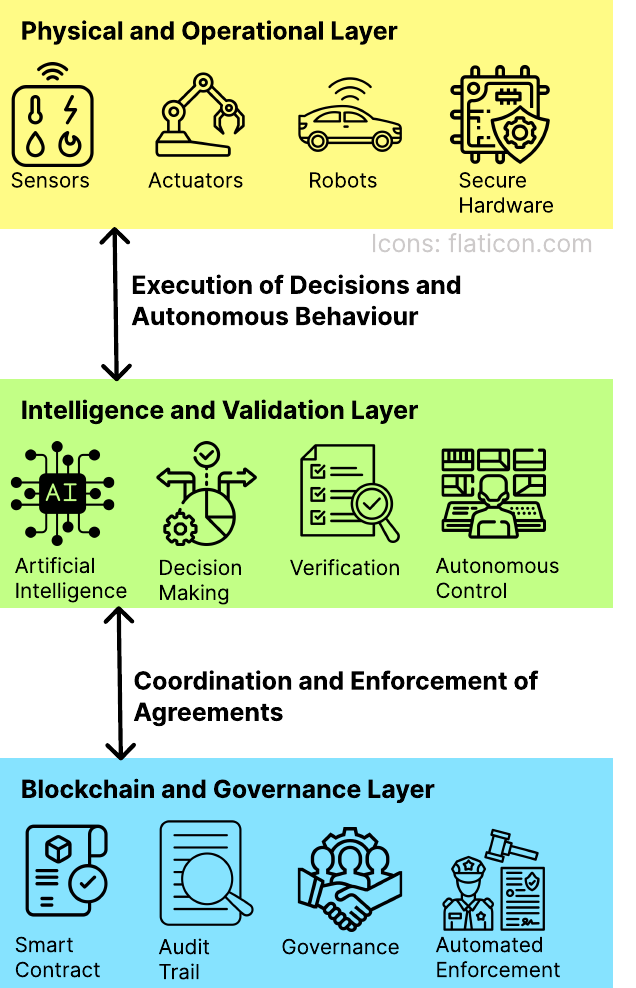}
    \caption{Layered Representation of a DAM}
    \label{fig:layered-dam}
\end{figure}

A DAM represents a machine that acts as an autonomous service provider, participating in a market and executing activities like economic transactions with minimal human intervention. In practical terms, this means that a DAM can manage RDWAs, as both digital assets (like cryptocurrency, data, or digital services access) and physical assets, without needing a person in the loop. For example, an autonomous vehicle functioning as a DAM could possess a digital wallet, automatically accept payments for rides, pay for its fuel or maintenance, and coordinate logistics with other machines – all according to pre-set AI logic and blockchain-based rules. Such a machine uses IoT sensors to perceive conditions (traffic, wear and tear), AI algorithms to make decisions (when to recharge or where to offer rides), and blockchain to securely handle identity, data, and payments. By combining these technologies, DAMs create a bridge between the physical world of machines and the digital world of decentralized networks, allowing devices to autonomously generate value and manage resources on behalf of their owners or stakeholders. This layered architecture of DAMs in illustrated in~\Cref{fig:layered-dam}. 

\subsection{Layered Architecture}
As a subset of DePIN, preserving part of its core mechanisms~\cite{lin2024decentralized, pillai2024sok, fan2023towards}, DAMs also operate through a multi-layered architecture:

\begin{itemize}
    \item \textbf{Physical/Operational Layer:} 
    This layer consists of IoT devices (e.g., sensors, actuators, robots) that collect data and execute actions in the physical world. In this layer, the integrity of data acquisition and device authentication is crucial. Secure hardware integration may aid in preventing data tampering and ensuring operational correctness~\cite{heiss2024towards,heiss2021trustworthy}.
    
    \item \textbf{Intelligence \& Validation Layer:} 
    In this layer, AI and ML algorithms process sensor data, facilitate decision-making, and validate operations off-chain. The AI controlling the DAM performs computing and real-time analytics, improving system responsiveness and optimizing resource allocation~\cite{ahmed2022blockchain}.%
    
    \item \textbf{Blockchain \& Governance Layer:} 
    The decentralized ledger records transactions and enforces rules via smart contracts, providing an immutable audit trail for machine interactions. This layer ensures trustless coordination among autonomous agents and supports automated governance mechanisms, replacing traditional centralized oversight~\cite{sedrati2022iot}.
\end{itemize}

\subsection{Trustworthy Validation and Verification}

Due to the decentralized nature of these interactions, a trustless mechanism is required for validating and verifying claims used in interactions with DAMs or those generated by the DAMs themselves. From this perspective, oracles serve as intermediaries between off-chain data and on-chain smart contracts, fetching, verifying, and relaying external information~\cite{heiss2019oracles}. This enables DAMs to respond to real-world events and maintain dynamic interactions with their environment. To achieve trustless verification, Verifiable Computing techniques, such as Trusted Execution Environments (TEEs) and Zero-Knowledge Proofs (ZKPs), are required. %

\textbf{Zero-Knowledge Proofs:} These cryptographic mechanisms enable a party to prove the validity of a statement without disclosing sensitive underlying data~\cite{christ2024sok}. This privacy-preserving approach is important for validating and verifying off-chain computations and sensor readings while maintaining the confidentiality of proprietary information. For example, zkSTARKs allow computations to generate proofs that can be audited in a trustless manner~\cite{ben2019scalable}. Additionally, ZKPs contribute to scalability by reducing the computational and storage burden on-chain, enabling more efficient transaction verification and data integrity checks~\cite{christ2024sok}.
    
\textbf{Trusted Execution Environments:} They establish a secure enclave within a device’s processor, ensuring that critical computations, such as data processing and cryptographic signing, remain isolated from potential system compromises. This hardware-based security is important for preserving the integrity of operations, particularly in secure oracle implementations~\cite{heiss2021trustworthy, schneider2022sok}. For instance, Intel TDX enables attestation quotes to be directly linked to a specific binary image, providing verifiable proof of the executed code~\cite{cheng2024intel}.

\textbf{Software Validation:} Another challenge is ensuring the trustworthiness of the software running in the DAMs, especially in trustless environments where both humans and machines must verify and validate security properties, operational features, or compliance measures. Software validation is key, as vulnerabilities can stem from external dependencies, unverified updates, or misconfigurations. Risks include software supply chain attacks~\cite{williams2024research}, such as compromised libraries, and operational negligence~\cite{catuogno2023secure}, including misconfigured circuits or faulty firmware updates, which threaten system integrity and trust. To mitigate these threats, methodologies like DevSecOps, VeriDevOps, and TrustOps show how to integrate security, verification, and trust mechanisms throughout the software life cycle~\cite{brito2024trustops}. These frameworks adopt a holistic approach, combining automation, continuous verification, and policy enforcement to enhance trustworthiness in decentralized and adversarial contexts.

\subsection{AI and IoT for self-managed RDWAs}

Real and Digital World Assets (RDWAs), nodes from the DePIN perspective, typically require managers to perform specific activities for their upkeep and operation~\cite{ford2022operational}. With the integration AI and IoT, DAMs can autonomously handle these activities, significantly reducing the need for human intervention. DAMs can assume various management tasks for RDWAs, including:

\begin{itemize}
    \item \textbf{Gathering and interpreting information:} Information related to the asset can be obtained through sensors. Information can be interpreted by AI algorithms, e.g., large language models or classification models~\cite{xu2024penetrative}.

    \item \textbf{Quantifying asset health:} The health of an asset is crucial for its proper operation. Health checks can be performed using existing models tailored to each asset, but AI can also enhance this process by enabling predictive maintenance strategies that help preserve the asset's functionality~\cite{shin2021ai}.

    \item \textbf{Coordinating:} Handling the course of actions with other service providers for refurbishment, replacement, inspection or testing in case it is required. This can be performed with NLP capabilities using the information gathered from the machine's sensors~\cite{tran2025multi}.

    \item \textbf{Operating:} Certain assets require operation to generate revenue, unlike passive or store-of-value assets, which may appreciate in value without requiring active operation. Some basic operations of these assets can already be automated, e.g., a vending machine or electrical vehicle (EV) charger, or be automated using AI, e.g., an autonomous vehicle~\cite{guo2022autonomous}.
    
\end{itemize}

\begin{figure}[h]
    \centering
    \includegraphics[width=0.8\columnwidth]{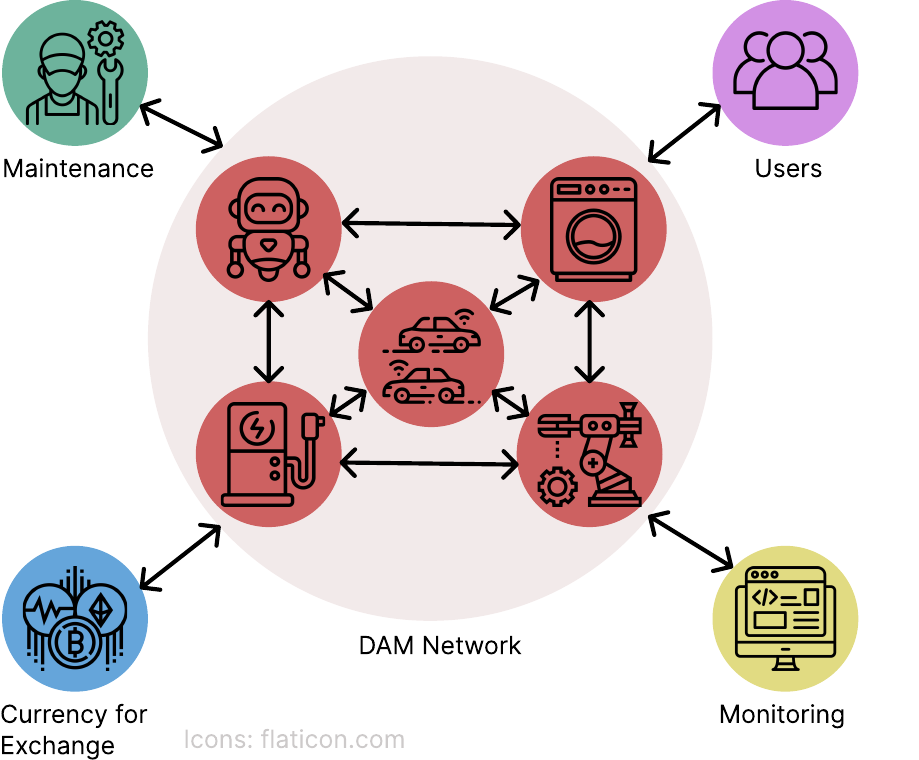}
    \caption{Interactions within a DAM Network, with internal M2M and external interactions}
    \label{fig:network-dam}
\end{figure}

Autonomous interactions within a DAM network can occur between DAMs or with external entities, as shown in~\Cref{fig:network-dam}. For instance, an EV might independently charge at a station, while a robotic assistant handles household tasks, all facilitated by blockchain for secure transactions and verifications. Additionally, DAMs enable self-organization and reorganization depending on the available infrastructure~\cite{picard2015multi}, allowing them to adapt their operations and network structure dynamically to optimize efficiency and resource use. This flexibility ensures that DAMs remain effective even as their environment or available resources change.

\subsection{Scalability and Interoperability}
As autonomous machine networks expand, ensuring scalability and integration across diverse systems becomes a critical consideration. Emerging blockchain scalability solutions can be explored to support the expansion of DAM and RDWA applications:
\begin{itemize}
    \item \textbf{Layer-2 Solutions:} 
    Techniques such as state channels, sidechains, and rollups process transactions outside of the main blockchain, reducing latency and increasing throughput. These solutions have been shown to significantly scale blockchain performance~\cite{xu2022l2chain}, and may become essential for supporting high-volume interactions, typical to IoT environments.
    
    \item \textbf{Cross-Chain Interoperability:} 
    Protocols enabling cross-chain communication (e.g., atomic swaps and relay chains) allow different blockchain networks to interact. This interoperability ensures that DAMs operating on different ledgers can share information and value without centralized coordination. Such factor is critical for applications spanning multiple sectors like energy management and robotics~\cite{fan2023towards}.
\end{itemize}

Overall, this baseline framework enables DAMs to autonomously bridge physical and digital domains, facilitating applications in decentralized economies. These applications may have broader socio-economic implications, which we examine in more depth in the next section.  %

\section{Socio-Economic Implications of DAMs}
\label{sec:Evaluation}

\begin{figure}[h]
    \centering
    \includegraphics[width=0.9\columnwidth]{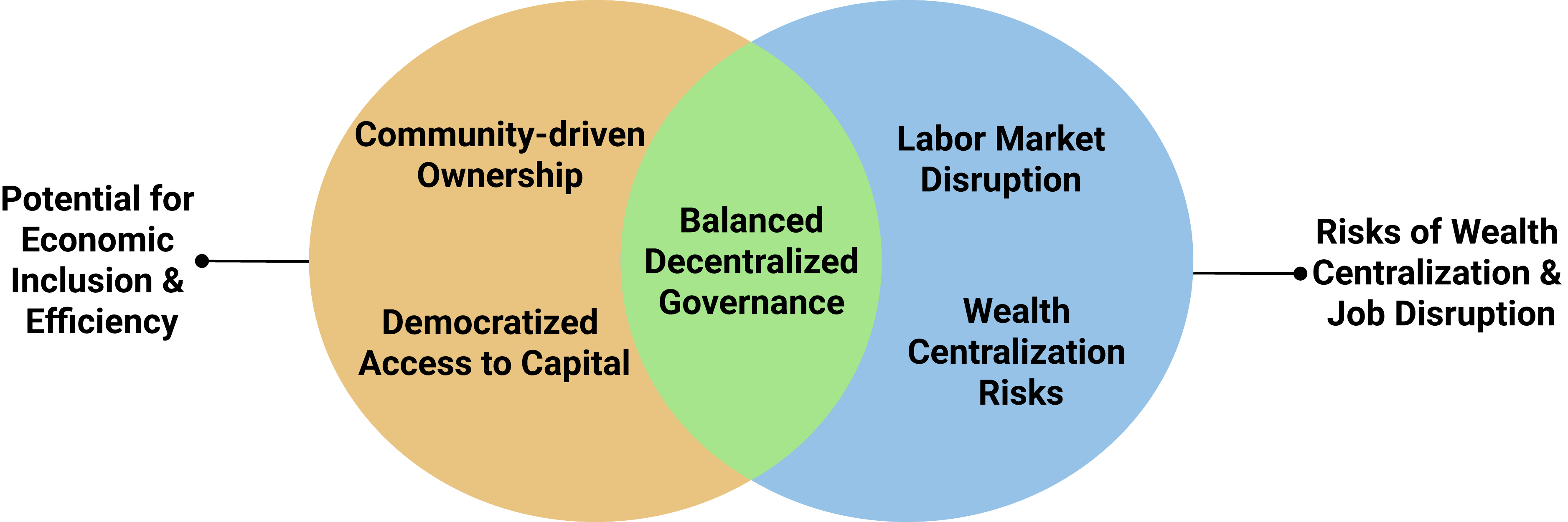}
    \caption{Venn Diagram illustrating the Spectrum of the Impact of DAMs}
    \label{fig:equitable-dams}
\end{figure}

The transformative potential of DAMs extends beyond technical innovation, promising also socio-economic impacts, as shown in~\Cref{fig:equitable-dams}. The automation of decision-making, financial transactions, and even entire operational workflows, may lead DAMs to redefine how wealth is generated and distributed. Their continuous, real-time operation has the potential to drive significant efficiency gains, eliminating intermediaries, reducing costs, and enabling direct peer-to-peer exchanges via DeFi protocols.%

On the one hand, DAMs facilitate a rethinking of ownership through tokenization. The representation of tangible RWAs (e.g., real estate, infrastructure, industrial equipment) as digital tokens enables fractional ownership, lowering entry barriers for individual investors and democratizing access to capital markets~\cite{baum2021tokenization}. Early DAO initiatives illustrate how community-driven models can transform traditional asset ownership into a shared, decentralized establishment~\cite{wang2019decentralized}. Potentially, this distributed ownership can lead to a more inclusive wealth creation process, where even small stakeholders benefit from the income generated by tokenized assets.

On the other hand, the widespread deployment of DAMs may accelerate automation to a degree that disrupts existing labor markets. It is projected by recent research~\cite{di2023future} that the accelerated pace of automation will demand retraining for 6 out of 10 jobs by 2027. If DAMs start extending automation into service and management sectors, there is a tangible risk that similar job losses could occur, raising important questions about the future role of human labor in an increasingly automated economy.

Moreover, while decentralized networks promise to distribute economic power more equitably, they also carry the risk of wealth concentration. Early adopters or those with significant technical expertise may accumulate a disproportionate share of tokens and mining rewards, thereby reinforcing or even exacerbating existing inequalities. Additionally, decentralized governance models, despite their potential to empower communities, may face challenges such as low voter participation and the outsized influence of whale stakeholders, potentially leading to decision-making that favors a small group over the broader community~\cite{sai2021characterizing}.

The rise of DAMs could reshape the insurance industry, a key pillar of social and economic stability. As DAMs autonomously manage assets, they shift risk dynamics, demanding new insurance models. Blockchain-based smart contracts can optimize underwriting, premium collection, and claims processing, reducing costs and enhancing transparency~\cite{brophy2020blockchain}. AI-driven risk assessments further refine pricing and policy customization, expanding accessibility~\cite{zarifis2023evaluating}. However, the decentralized nature of DAMs complicates liability assignment in system failures or accidents~\cite{saenz2023autonomous}. Developing adaptive regulations and insurance solutions will be important to mitigating socio-economic risks while fostering financial inclusion.

In summary, we see DAMs as a convergence of robotics, AI, and blockchain that challenge conventional business models by redistributing ownership and decision power. While they offer the promise of increased efficiency and democratized asset ownership, we acknowledge that their socio-economic benefits may depend on carefully designed tokenomics and governance structures. Ensuring that these systems foster inclusive growth rather than replicating traditional power imbalances remains a critical challenge as society navigates this new digital-physical frontier.

\section{Conclusion}
\label{sec:Conclusion}
In this paper, we introduce Decentralized Autonomous Machines (DAMs) as a transformative integration of AI, blockchain, and IoT. We have explored how the combination of these technologies enables emerging capabilities: machines that can autonomously manage RDWAs, execute financial transactions, and participate in decentralized marketplaces without human intervention. This shift from centralized to decentralized control mechanisms addresses fundamental inefficiencies in current systems by reducing reliance on intermediaries, increasing operational transparency, and enhancing economic inclusion through fractional ownership models.

However, we recognize that the widespread adoption of DAMs presents substantial challenges that must be addressed. From a technical perspective, scalability limitations, interoperability barriers, and security vulnerabilities require ongoing research and development. From a socio-economic standpoint, the potential displacement of labor, risks of wealth concentration among early adopters, and governance complexities necessitate careful consideration and proactive policy responses.

The ultimate impact of DAMs will depend on how we navigate these challenges. Effective implementation will require governance frameworks, equitable tokenomics models, regulatory approaches, education and workforce transition programs. If these considerations are addressed, we believe that DAMs may have the potential to democratize access to economic infrastructure, optimize resource allocation, and create new forms of value that bridge the physical and digital realms.

\section*{Acknowledgements}{Funded by the European Union (TEADAL, 101070186). Views and opinions expressed are, however, those of the author(s) only and do not necessarily reflect those of the European Union. Neither the European Union nor the granting authority can be held responsible for them.}




\end{document}